\documentclass[twocolumn,prl,aps,epsf,showpacs]{revtex4}

\usepackage{amssymb,epsf}
\usepackage{graphicx}
\usepackage{epstopdf}

\begin{document}

%\title{Nuclear Spin Relaxation by Electron Localization in the Quantum Hall Effect}
%\title{Evidence for Skyrmion Crystallization from NMR Relaxation Experiments}

\author{C.R. Dean$^{1,\star}$, B.A. Piot$^{1,\star}$, P. Hayden$^{2}$, S. Das Sarma$^{3}$, G. Gervais$^{1}$, L.N. Pfeiffer$^{4}$ and K.W. West$^{4}$ }

\address{$^{1}$Department of Physics, McGill University, Montreal, H3A 2T8, CANADA}
\address{$^{2}$School of Computer Science, McGill University, Montreal, H3A 2A7, CANADA}
\address{$^{3}$Condensed Matter Theory Center, Department of Physics, University of Maryland, College Park, MD 20742 USA}
%\address{$^{3}$Department of Physics and Department of Applied Physics, Columbia University, New York, NY 10027 USA}
\address{$^{4}$Bell Laboratories, Lucent Technology, Murray Hill, NJ 07974 USA}

%\usepackage{graphicx}
%\usepackage{dcolumn}
%\usepackage{bm}

%%%%%%%%%%%%%%%%%%%%%%%%%%%%

\title{Intrinsic Gap of the $\nu=5/2$ Fractional Quantum Hall State}

\begin{abstract}

The fractional quantum Hall effect is observed at low magnetic field, in a regime where the cyclotron energy is smaller than the Coulomb interaction energy. The $\nu=\frac{5}{2}$ excitation gap is measured to be 262$\pm$15~mK at $\sim$2.6~T, in good agreement with previous measurements performed on samples with similar mobility, but with electronic density larger by a factor of two.  The role of disorder on the $\nu=\frac{5}{2}$ gap is examined. Comparison between experiment and theory indicates that a large discrepancy remains for the intrinsic gap extrapolated from the infinite mobility (zero disorder) limit. In contrast, no such large discrepancy is found for the $\nu=\frac{1}{3}$ Laughlin state. The observation of the $\frac{5}{2}$ state in the low-field regime implies  that inclusion of non-perturbative Landau level mixing may be necessary to better 
understand the energetics of half-filled fractional quantum Hall liquids.

\end{abstract}
 \maketitle

Since the discovery of the fractional quantum Hall effect (FQHE), understanding the role played by electron-electron interactions has been the source of major breakthroughs in our understanding of strongly interacting two-dimensional electron gases (2DEGs). Chief among these is the composite fermion picture of the incompressible FQH liquid \cite{Jain89,Halperin93},  extremely successful at explaining both the complete series of observed FQH states in the first Landau level (FLL), and the absence of such a liquid at precisely half-filling.  In the second Landau level (SLL), however, the situation is more complex where experiments have shown, unambiguously, exact quantization of the Hall resistance at filling factor $\nu=\frac{5}{2}$ \cite{Willett:PRB:1987,Pan:PRL:83(17):1999} and $\nu=\frac{7}{2}$. 
In 1991, Moore and Read \cite{MooreRead:NuclPhysB:1991}  proposed an elegant many-body wave function to explain this phenomenon that described the $\frac{5}{2}$ FQH state as a `condensation process' of composite fermions. In recent years, this Moore-Read `Pfaffian' state has received considerable interest owing to built-in quantum statistics that are now predicted to be {\it non-abelian}.  The non-abelian composite particles that comprise the $\nu=\frac{5}{2}$ FQH state underlie a paradigm for fault-tolerant topological quantum computation first proposed by Kitaev\cite{Kitaev}  and recently exploited by Das Sarma, Freedman and Nayak\cite{Nayak}. Yet, in spite of these many recent theoretical advances, an unequivocal experimental verification of the Moore-Read description is still missing.  Furthermore, continued discrepancies between experiment and theory, such as the large difference between the measured and calculated activation energy gap, remain problematic.

In an effort to better understand electron-electron interaction at half filling, we present in this work a detailed analysis of the $\nu=\frac{5}{2}$ state for a sample with, to our knowledge, the lowest electron density reported to date (by nearly a factor of two).  This allows the study of the FQHE in a regime where the cyclotron energy is smaller than the Coulomb interaction energy. We compare the measured energy gap with neighbouring FQH states in the SLL, and discuss these results in the context of previous studies allowing us to deduce the intrinsic gap in the zero-disorder limit.  Our analysis  shows that large discrepancies remain between theory based on a Moore-Read Pfaffian state and experiment at $\nu=\frac{5}{2}$ that cannot be attributed to disorder alone. In contrast,  a similar analysis for the $\nu=\frac {1}{3}$ Laughlin state shows much better agreement with current models.

The sample used in this study was a 40 nm wide, modulation-doped, GaAs/AlGaAs quantum well, with a measured density of $1.6(1)\times10^{11}~\text{cm}^{-2}$ and mobility of $14(2)\times10^{6}~\text{cm}^{2}/\text{V$\cdot$s}$.  The sample 
was cooled in a dilution fridge enclosed inside a shielded room, with a base temperature of $\sim$16 mK in continuous mode, and equipped with a 9 Tesla magnet.  Treatment with a red LED was used during the cooldown.  \textit{In situ} powder filters and RC filters were used on the sample leads to ensure efficient cooling of the 2DEG.  Temperatures were monitored with a RuO resistive thermometer, and a CMN magnetization thermometer, both calibrated with superconducting fixed points.  Transport measurements were performed using a standard lock-in technique at $\sim$6.5~Hz.  

\begin{figure}[t]
	\begin{center}
		\includegraphics[angle=0,clip]{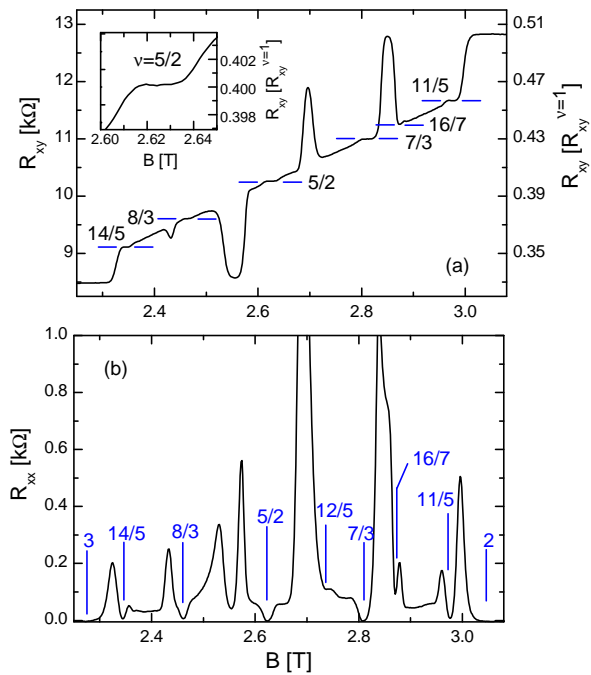}
		\caption{(a) Hall resistance and (b) corresponding magnetoresistance in the second Landau level of our low density, high mobility 2DEG (T$\sim$20~mK).}
		\label{Fig1}
	\end{center}
\end{figure}

Fig.~\ref{Fig1} shows the magnetoresistance (R$_{xx}$) and corresponding Hall resistance (R$_{xy}$), taken around $\nu=\frac{5}{2}$ in the SLL at $\sim$20~mK.  A vanishingly small magnetoresistance is observed at $\nu=\frac{5}{2}$, which, together with a wide plateau in the corresponding Hall trace, indicates the $\frac{5}{2}$ state is exceptionally well formed.  The unambiguous $\frac{5}{2}$ state observed here, occurring at $\sim$2.63~T, represents to our knowledge the lowest magnetic field observation of the $\frac{5}{2}$ to date \cite{Willett:PRB:1987,Pan:PRL:83(17):1999, Eisenstein:PRL:1988, Gammel:PRB:1988,Sajoto:PRB:1990, Eisenstein:SurfSci:1990,Pan:PRL:83:1999,Pan:SolStatComm:2001, Eisenstein:PRL:2002, Xia:PRL:2004, Csathy:PRL:2005, Choi:arXiv:2007, Willet:APL:91, Miller:Nature:2007, Pan:arXiv:2008}. Strong FQHE minima are also observed at $\nu=\frac{14}{5},\frac{8}{3},\frac{7}{3},\frac{16}{7}$, and $\frac{11}{5}$, each of which exhibit plateaus in R$_{xy}$.  A hint of an emerging R$_{xx}$ minimum can also be seen at $\nu=\frac{12}{5}$. The four reentrant phases observed in the Hall trace on either side of the $\frac{5}{2}$ plateau (two peaks tending towards R$_{xy}$=h/3e$^{2}$ and two tending towards R$_{xy}$=h/2e$^{2}$) together with the observation of the $\nu=\frac{16}{7}$ state, and the emerging minimum at $\nu=\frac{12}{5}$, are all signatures of an extremely high quality sample \cite{Pan:PRL:83(17):1999, Eisenstein:PRL:2002, Xia:PRL:2004, Choi:arXiv:2007}.  The deep R$_{xx}$ minima appearing in the reentrant insulating phase at $\sim$2.55~T (Fig.~\ref{Fig1}b) is similar to that observed elsewhere upon lowering the electronic temperature to a regime where the reentrant state is fully formed \cite{Miller:Nature:2007, Pan:arXiv:2008, Xia:PRL:2004}.  

\begin{figure}[t]
	\begin{center}
		\includegraphics[angle=0,clip]{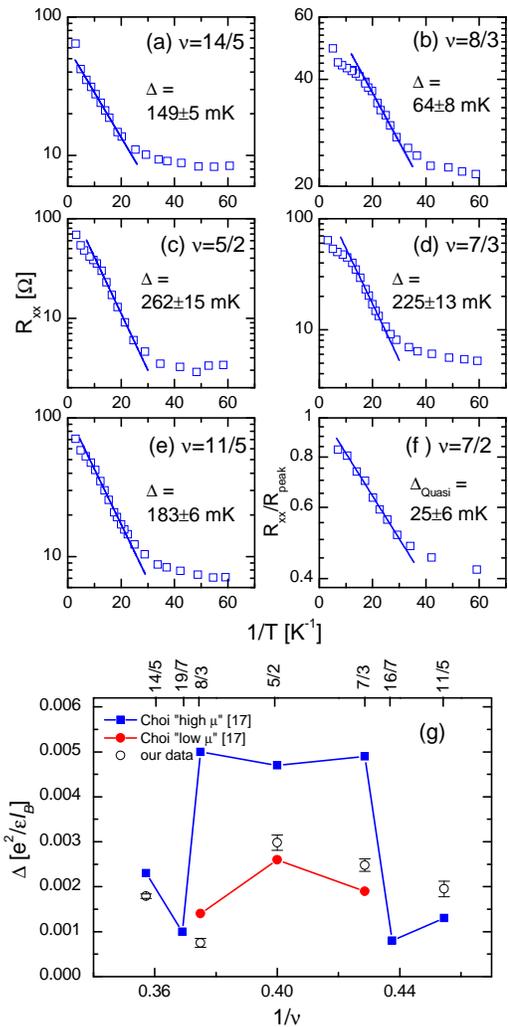}
		\caption{(a)-(e) Arrhenius plot showing activated temperature behaviour in the second Landau level.  (f) Quasi-gap measurement for $\nu=\frac{7}{2}$ (see text). (g) Energy gaps for the FQH states, plotted in Coulomb energy units.  Open circles are our data.  Solid squares and circles, respectively, refer to the ``high mobility'' and ``low mobility'' samples in reference \cite{Choi:arXiv:2007}.}
		\label{Fig2}
	\end{center}
\end{figure}

The temperature dependence of the FQHE minima are shown in Fig.~\ref{Fig2}a-e,
with all data acquired at fixed magnetic field in order to avoid 
heating effects caused by varying fields.
 The corresponding energy gaps were determined by linear fits to the thermally activated transport region, where the resistance is given by the equation R$_{xx}\propto e^{-\Delta/2k_{B}T}$.  The gap error quoted on each plot was estimated from the goodness of the linear fit.  Examination of weakly formed FQHE states under single shot of the dilution fridge down to $\sim$9~mK indicated the electrons continued to cool, suggesting that the low temperature tail-off observed in the data of Fig.~\ref{Fig2} does not reflect a saturation in the electronic temperature.  Instead, it may indicate a transition from activated conduction to hopping conduction \cite{Boebinger:PRL:1985}, and/or could result from the energy dependent Landau level broadening due to disorder \cite{Usher:PRB:1990}. In the $\frac{7}{3}$, $\frac{5}{2}$, and $\frac{8}{3}$ states (Figs.~\ref{Fig2}b-d), there is a deviation from activated behaviour at high temperature whose onset temperature scales with the corresponding gap value.  Likely, this results from $k_{B}T$ approaching the gap value.  However, the same deviation is not observed in the $\frac{11}{5}$ and $\frac{14}{5}$ FQH states, which have lower measured gaps than the $\frac{7}{3}$ and $\frac{5}{2}$ states.  Interestingly, recent work has suggested the $\frac{11}{5}$ and $\frac{14}{5}$ to be Laughlin states, while the  $\frac{7}{3}$ and $\frac{8}{3}$ are proposed to be non-Laughlin \cite{Choi:arXiv:2007, Bonderson:arXiv:2007, Simion:arXiv:2007}.

We also observed a FQH state at $\nu=\frac{7}{2}$, the electron-hole conjugate of $\nu=\frac{5}{2}$,  appearing at a magnetic field less than 2~T.  However, due to a competition between the weakly formed $\frac{7}{2}$ and rapidly emergent neighbouring reentrant states, the R$_{xx}$ minimum did not fall significantly with temperature near base. The ``quasi-gap'' was therefore determined by measuring the depth of the $\frac{7}{2}$ minima with respect to the average resistance of the two neighbouring peaks (R$_{peak}$)\cite{Eisenstein:PRL:1988, Pan:SolStatComm:2001, Gammel:PRB:1988}.  The resulting Arrhenius plot, which clearly shows activated behaviour (Fig.~\ref{Fig2}f) gives an estimate for the $\frac{7}{2}$ gap value of $\sim$25~mK.

In Fig.~\ref{Fig2}g, the gap values are plotted in Coulomb energy units, e$^{2}$/$\epsilon l_{B}$, where $l_{B}=\sqrt{\hbar/eB}$ is the magnetic length, and $\epsilon=12.9$ is the dielectric constant.  Results from recent gap measurements in the SLL by Choi \textit{et al.} are also shown for comparison \cite{Choi:arXiv:2007}.  The excellent agreement between our data set and that of the  Choi \textit{et al.} `low mobility' sample ($\mu_{B}=10.5\times10^{6}\text{cm}^{2}/\text{V$\cdot$ s}$) is surprising given the factor of two difference in electron densities between our sample ($1.6\times10^{11}\text{cm}^{-2}$) and theirs ($2.8\times10^{11}\text{cm}^{-2}$ and $3.2\times10^{11}\text{cm}^{-2}$ for the ``low mobility'' and ``high mobility'' respectively).  Simple dimensional considerations imply that the interaction energy, and hence the FQH gap, should scale as $\sqrt{B}$, which would predict a $\sim$40\% enhancement in the gap between the low density (ours) and the high density (Choi {\it et al.}) samples.  %Our finding that the gap is almost the same for the two samples suggests that disorder and/or possibly non-perturbative Landau level coupling play a role in determining the 5/2 FQH gap in a way not yet understood theoretically.
Our finding that the gap is almost the same for the two samples with similar mobility (independent of density), while significantly enhanced in samples with higher mobility (Choi \textit{et al.} ``high mobility'') indicates that disorder more strongly affects the gap than the applied magnetic field.  Furthermore, the similar gap value measured in a low magnetic field where the cyclotron energy is reduced compared to the Coulomb interaction suggests non-perturbative Landau level coupling may affect the $\nu=\frac{5}{2}$ FQH gap in a way not yet understood theoretically.

%\begin{table}[t]
%\centering \caption{Measured gap energies for the fractions shown in Fig.~\ref{Fig2} expressed in both mK and Coulomb %energy units.}
%\label{Table1}
%\begin{tabular}{l c c c c c c}
%\hline\hline
%{}& \multicolumn{6}{c}{fraction ($\nu$)}\\
%{} &{7/2}&{14/5}&{8/3}&{5/2}&{7/3}&{11/5}\\ [0.5ex]
%\hline 
%\vspace{-0.1in}
%\\
% {$\Delta$ [mK]} &{25$\pm$5}&{149$\pm$5}&{64$\pm$8}&{262$\pm$15}&{225$\pm$13}&{183$\pm$16} \\[0.5ex]
% {$\Delta$ [$\frac{e^{2}}{\epsilon l}$]} &{ 0.0004 }&{ 0.0018 }&{ 0.0008 }&{ 0.0030 }&{ 0.0025 }&{ 0.0020 } \\[1ex]
%\hline\hline
%\end{tabular}
%\end{table}

\begin{figure}[t]
	\begin{center}
		\includegraphics[width=0.8\linewidth,angle=0,clip]{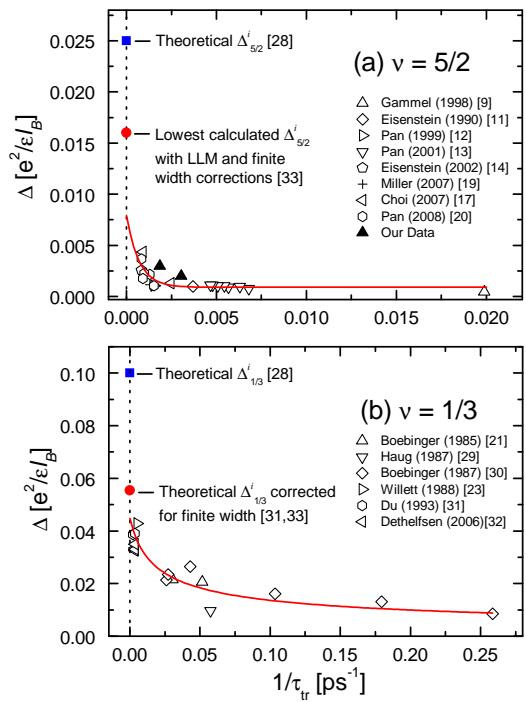}
		\caption{(a)  Comparison of the measured $\nu=\frac{5}{2}$ gap energy with values found in literature (open symbols). Solid triangles represent our data.  Solid square is the theoretically calculated intrinsic gap energy \protect\cite{Morf:PRB:2002,Feiguin:arXiv:2007} and the solid circle includes corrections for Landau level mixing, and finite width. (b) Same plot as in (a), but for $\nu=\frac{1}{3}$ energy gap values reported in the literature.} 
		\label{Fig3}
	\end{center}
\end{figure}

In Fig.~\ref{Fig3}a, we show a plot of all the $\frac{5}{2}$ gap values found in the literature versus the inverse transport lifetime, $\tau_{tr}^{-1}$, deduced from the reported mobilities \cite{Gammel:PRB:1988, Pan:SolStatComm:2001, Eisenstein:SurfSci:1990, Pan:PRL:83(17):1999, Eisenstein:PRL:2002, Miller:Nature:2007, Choi:arXiv:2007, Pan:arXiv:2008}.  In spite of the large spread in the $\frac{5}{2} $ data, owing to wide ranging differences in sample parameters, {\it i.e.} dopant, well width, etc.,  a clearly discernible trend (indicated by the solid curve as a guide-to-the-eye) is observed pointing towards a disorder-free intrinsic gap value in the
range of $\Delta^{i}_{5/2}\sim$ 0.005-0.010~e$^{2}$/$\epsilon l_{B}$. This estimate is in good agreement with a similar extrapolation reported very recently by Pan {\it et al.} \cite{Pan:arXiv:2008,SimilarPlots}.  Moreover, examination of the low field Shubnikov de Haas oscillations gave the level broadening, $\Gamma$, in our sample to be $\Gamma=0.168\pm0.040$~K.  This gives a  direct experimental estimate for the intrinsic gap, $\Delta^{i}=\Delta^{exp} + \Gamma$, of $\sim$0.005~e$^{2}$/$\epsilon l_{B}$, also in good agreement with the extrapolated intrinsic gap value in Fig.~\ref{Fig3}a.  Importantly, the experimentally measured intrinsic gap inferred from this data remains well below (by a factor of three to five)  the theoretically estimated intrinsic gap value for a Moore-Read type Pfaffian wave function,  ($\sim$0.025~e$^{2}$/$\epsilon l_{B}$) \cite{Morf:PRB:2002, Feiguin:arXiv:2007}. 

\begin{figure}[t]
	\begin{center}
		\includegraphics[width=0.75\linewidth,angle=0,clip]{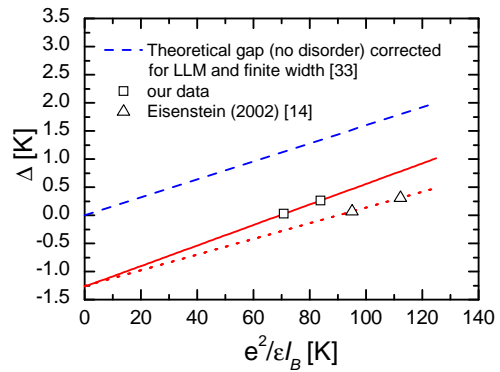}
		\caption{Intrinsic gap energy at $\nu=\frac{5}{2}$ from particle-hole paired states \protect\cite{Morf:PRL:2003} (see text). Dashed line indicates a theoretical gap with no disorder.  Open squares are the gaps for the  $\frac{7}{2}$ and $\frac{5}{2}$ FQH states.  Open triangles are data taken from Ref. \protect\cite{Eisenstein:PRL:2002}, for comparison. Solid and dotted lines
		are linear fit to the data.}
		\label{Fig4}
	\end{center}
\end{figure}

For comparison, a similar plot  for the measured gap values of the $\frac{1}{3}$ Laughlin  state is shown in Fig.~\ref{Fig3}b, \cite{Boebinger:PRL:1985, Haug:PRB:1987, Boebinger:PRB:1987, Willett:PRB:1988, Du:PRL:1993, Dethlefsen:PRB:2006}.  In contrast to the $\nu=\frac{5}{2}$ FQH state, the intrinsic gap determined at $\nu=\frac{1}{3}$ with our procedure ($\Delta ^{i}_{1/3}\sim0.045$~e$^{2}$/$\epsilon l_{B}$) is in good  agreement with theory ($\sim0.055~\text{e}^{2}/\epsilon l_{B}$) \cite{Morf:PRB:2002, Du:PRL:1993}.  Morf \textit{et al.} proposed that since the disorder-induced Landau level broadening is expected to be roughly equal for FQH states corresponding to particle-hole conjugate pairs, then plotting the correspondig gap values as a function of Coulomb energy directly gives a measure for the intrinsic gap (slope of a fitted line to this data) \cite{Morf:PRL:2003}.
%Morf \textit{et al.} proposed that the level broadening, $\Gamma$, is filling fraction dependent, so that the experimentally measured gap is described by $\Delta^{exp}(\nu)=\Delta^{i}(\nu) - \Gamma(\nu)$, where $\Delta^{i}(\nu)=\delta(\nu)E_{c}$ is the intrinsic gap with $\delta(\nu)$ termed the intrinsic gap parameter.  However, $\Gamma(\nu)$ is expected to be roughly equal for FQH states corresponding to particle-hole conjugate pairs.  Plotting the experimental gaps for conjugate pairs as a function of their corresponding Coulomb energy ($E_{c}=e^{2}/\epsilon l_{B}$), therefore gives a linear relationship where the slope ($\delta(\nu)$) provides a direct measure of the intrinsic gap, and the intercept, $\Gamma(\nu)$, the level broadening.  
Fig. \ref{Fig4} shows the $\frac{5}{2}$ and $\frac{7}{2}$ gap values obtained in our low electron density sample (open squares) together with those from Ref. \cite{Eisenstein:PRL:2002} (open triangles).  The dashed line shows the predicted trend for a disorder free gap.  The slope extracted from a linear fit gives the intrinsic gap for our sample to be $\sim$0.018~e$^{2}$/$\epsilon l_{B}$, which is in remarkable agreement with the data from ref. \cite{Eisenstein:PRL:2002} ($\sim$0.014) and the theoretical value ($\sim$0.016) corrected for the sample parameters specified in ref. \cite{Eisenstein:PRL:2002,Morf:PRL:2003}.  This however disagrees with the intrinsic gap estimated both from our sample, and from the  extrapolation towards the infinite mobility limit.  Furthermore,  the Landau Level broadening deduced from Fig.~\ref{Fig4} implies a rather large value ($\sim$1.25~K) that is an order of magnitude larger than the value determined experimentally from the SdH oscillations ($\sim$0.17~K).
 
It is instructive to consider the energetics of our low-density $\frac{5}{2}$ FQH state.
At the observed field of 2.62~T the cyclotron energy is 52~K, the interaction energy is 81~K, and the Zeeman energy
(assuming the GaAs band g-factor) is 0.75~K. The level broadening in our sample was measured to be $\sim$0.17~K and the  mobility broadening $\sim$0.006~K \cite{DasSarma:PRB:1985}.  Also important is the suppression of the ideal 
two-dimensional  FQH excitation gap due to the finite width $d=40$~nm of our quasi-2D square well sample. For our $\frac{5}{2}$ FQH gap, this is only about 15\% (using $d/l_{B}=2.5$ in \cite{Morf:PRB:2002}, where $l_{B}=16$~nm at $\sim$2.6~T). Taking all of these energies into account we conclude: $i)$ our measured gap value of 0.262~K is at least a factor of 5 lower than the ideal 2D theoretical $\frac{5}{2}$ excitation gap ($\sim$2~K at 2.6~T), even if the theoretical gap is corrected for finite width and level broadening suppression ($\sim$1.5~K);  $ii)$ the cyclotron gap, {\it i.e.} the Landau level separation,  is smaller than the interaction energy in our system, suggesting considerable non-perturbative inter-Landau level coupling, which has not so far been included in the theory, may be important in understanding the $\frac{5}{2}$ FQH state; {\it iii)} the Zeeman energy 
at 2.6~T is extremely small compared with the Coulomb energy, so
 the observation of a strong $\frac{5}{2}$ gap at this field might suggest that the $\frac{5}{2}$ FQHE is spin-unpolarized.  However, the $\frac{5}{2}$ FQHE has been observed in magnetic fields as large as 12T\cite{Pan:SolStatComm:2001}, {\it i.e.} with an increase in Zeeman energy by a factor of five, where the system is most likely spin-polarized, without affecting much the $\frac{5}{2}$ gap.  Therefore, unless a quantum phase transition occurs between a low-field ($\sim$2.6~T) spin- unpolarized state, and a high-field ($\sim$12~T) spin-polarized $\frac{5}{2}$ FQH state, our experiment rather points towards a spin-polarized state at $\nu=\frac{5}{2}$ even in the zero-field limit, consistent with a Moore-Read pfaffian wavefunction, the leading candidate for the $\frac{5}{2}$ FQHE.

In conclusion, the $\frac{5}{2}$ energy gap was measured for a sample with an electron density nearly twice smaller than previously observed, and was found to be comparable to samples with higher densities, and similar mobilities.  Extrapolating the experimentally measured energy gap values to zero disorder yields an estimate for the intrinsic gap which remains well below the theoretical value.  By contrast, a similar extrapolation for the $\frac{1}{3}$ Laughlin state  is in much better agreement with theory. Our study suggests that the large discrepancies observed between theory and experiment at $\nu=\frac{5}{2}$ cannot simply be attributed to disorder,
but rather may indicate that our knowledge of electron-electron interactions for the $\nu=\frac{5}{2}$ FQH state remains incomplete.  Based on the fact that the Coulomb interaction energy scale for our low density $\frac{5}{2}$ FQH
state is larger than the cyclotron energy, we speculate that the non-perturbative aspects of Landau level mixing (as well as disorder), not considered in the theoretical literature, may play an important role in the understanding of the enigmatic $\frac{5}{2}$ FQH state.

This work has been supported by the Natural Sciences and Engineering Research Council of Canada (NSERC), the Canada Fund for Innovation (CFI), the Canadian Institute for Advanced Research (CIFAR), the Canada Research Chairs program, MITACS, 
QuantumWorks and FQRNT (Qu\'ebec). Two of the authors (G.G. and P.H.) acknowledge receipt of support from the Alfred P. Sloan Foundation through their fellowship program.

{\it Special Note}: Both authors marked with a star (*) contributed evenly to this work.

\end{document}